\documentclass[12pt,reqno,a4paper,twoside]{article}
\usepackage{amsmath,amsthm,amstext,amscd,amssymb,euscript}
\usepackage{epsf}
\renewcommand{\phi}{\varphi}

\def\1{ {\mathit{1} \!\!\>\!\! I} }

\makeatletter
\@addtoreset{equation}{section}
\makeatother

\newtheorem{ittheorem}{Theorem}
\newtheorem{itlemma}{Lemma}
\newtheorem{itproposition}{Proposition}
\newtheorem{itdefinition}{Definition}
\newtheorem{itremark}{Remark}

\newenvironment{theorem}{\addtocounter{equation}{1}
\begin{ittheorem}}{\end{ittheorem}}

\newenvironment{lemma}{\addtocounter{equation}{1}
\begin{itlemma}}{\end{itlemma}}

\newenvironment{proposition}{\addtocounter{equation}{1}
\begin{itproposition}}{\end{itproposition}}

\newenvironment{definition}{\addtocounter{equation}{1}
\begin{itdefinition}}{\end{itdefinition}}

\newenvironment{remark}{\addtocounter{equation}{1}
\begin{itremark}}{\end{itremark}}

\newcommand{\beq}{\begin{eqnarray}}
\newcommand{\eeq}{\end{eqnarray}}

\newcommand{\be}{\begin{equation}}
\newcommand{\ee}{\end{equation}}

\newcommand{\bl}{\begin{lemma}}
\newcommand{\el}{\end{lemma}}

\newcommand{\br}{\begin{remark}}
\newcommand{\er}{\end{remark}}

\newcommand{\bt}{\begin{theorem}}
\newcommand{\et}{\end{theorem}}

\newcommand{\bd}{\begin{definition}}
\newcommand{\ed}{\end{definition}}

\newcommand{\bp}{\begin{proposition}}
\newcommand{\ep}{\end{proposition}}

\newcommand{\bc}{\begin{corollary}}
\newcommand{\ec}{\end{corollary}}

\newcommand{\bpr}{\begin{proof}}
\newcommand{\epr}{\end{proof}}

\newcommand{\bi}{\begin{itemize}}
\newcommand{\ei}{\end{itemize}}

\newcommand{\ben}{\begin{enumerate}}
\newcommand{\een}{\end{enumerate}}

\newcommand{\R}{\mathbb R}
\newcommand{\N}{\mathbb N}

\newcommand{\E}{\mathbb E}

\newcommand{\pee}{\ensuremath{\mathbb{P}}}

\newcommand{\ce}{\ensuremath{\mathcal{C}}}

\newcommand{\loc}{\ensuremath{\mathcal{L}}}

\newcommand{\si}{\ensuremath{\sigma}}

\newcommand{\epsi}{\ensuremath{\epsilon}}

\def\now{
\ifnum\time<60
          12:\ifnum\time<10 0\fi\number\time am
          \else
            \ifnum\time>719\chardef\a=`p\else\chardef\a=`a\fi
          \hour=\time
          \minute=\time
          \divide\hour by 60 
          \ifnum\hour>12\advance\hour by -12\advance\minute by-720 \fi
          \number\hour:%
          \multiply\hour by 60 
          \advance\minute by -\hour
          \ifnum\minute<10 0\fi\number\minute\a m\fi}
\newcount\hour
\newcount\minute
\numberwithin{equation}{section}         

\theoremstyle{remark}

\newcommand{\caG}{{\mathcal G}}

\newcommand{\bix}{\vec{x}}
\newcommand{\muT}{\mu_{T_L,T_R}}

\begin{document}

\title{{\bf Duality and exact correlations for a model of
heat conduction}}

\author{
Cristian Giardin\`a
\footnote{Department of Mathematics and Computer Science, 
Eindhoven University, P.O. Box 513 - 5600 MB Eindhoven, The Netherlands,
{\em c.giardina@tue.nl}}\\
Jorge Kurchan
\footnote{CNRS-ESPCI, rue Vauquelin 10, 75231 Paris, France,
{\em jorge@pmmh.espci.fr}}\\
Frank Redig
\footnote{Mathematisch Instituut Universiteit Leiden,
Niels Bohrweg 1, 2333 CA Leiden, The Netherlands,
{\em redig@math.leidenuniv.nl}}
}

\maketitle
\begin{abstract}
  We study a model of heat conduction with stochastic diffusion of energy.
 We obtain a  dual particle process which describes the evolution of
all the correlation functions. An
 exact expression for the covariance of the energy exhibits
  long-range correlations
in the presence of a current.
We discuss the formal connection
 of this model with the simple symmetric exclusion process.
\end{abstract}

\vspace{1.cm}

\section{Introduction}

Simple systems of particles on a lattice
have received considerable attention in the last years, as they
are a testing ground for exploring the properties of far from equilibrium
states, for which at present no general theory is available.

Amongst the most studied models are the so called exclusion processes
in which particles diffuse amongst empty sites on
a lattice. The problem has been attacked with techniques of
 statistical mechanics and probability, and many analytic results are available
\cite{lig,stinch,schutz-review}, in particular for the
simple symmetric exclusion process (SEP)

One is also interested in the transport of continuous quantities,
especially energy.
Kipnis, Marchioro and Presutti (KMP) \cite{KMP} have
 introduced a model of energy transport in which
the energy of neighboring sites is randomly redistributed.
This model has also been thoroughly studied, although analytic solutions - in particular for
the steady-state correlation functions - are
harder to obtain than in the SEP.
 Both the KMP and the SEP have an interesting (and rather exceptional) feature:
the evolution of the $K$-point correlation functions can be exactly mapped
 onto a diffusion equation for $K$ particles.
This so-called `duality property' \cite{lig} is a powerful tool,
which in the study of the SEP yields the complete ergodic theory (see \cite{lig}).

Despite the resemblance between energy and particle transport,
studying both models is not a redundant exercise, as
the SEP and the KMP models also show intriguing differences \cite{BGL},
mainly in their large-deviation functions.
Perhaps the most striking is the fact that in the implementation
of an `additivity principle' of the large deviation functions of two
subsystems, the density at the interface has to be maximized in one case,
and minimized on the other, in order to obtain the correct result \cite{DLS}.

In this paper we shall study a family of models of energy transport
which, we shall argue, are the natural counterpart of the SEP models.
In each site $i$ of the lattice there is a free particle with momentum
$x_i$. Between any two neighboring sites $(i,i+1)$ and for any small time interval
there is a random exchange  of momentum that leaves
$ \{ x_i^2 +  x_{i+1}^2 \}$ invariant.
The same transport terms  were already considered  in models of
wave propagation in  random media \cite{Mathur},\cite{Mathur1}. They later appeared
as the high-energy limit of a chain with deterministic dynamics \cite{GK},
and were also used as a stochastic perturbation (mimicking nonlinearities)
of an oscillator chain \cite{olla},\cite{olla2}.

In the present paper we  construct for our energy diffusion model a dual process that expresses
the evolution of
the $K$-point correlation functions of the kinetic energies
in terms of a process of $K$ interacting random walkers.
We also give a closed expression for the stationary covariance
$\langle x_i^2; x_j^2 \rangle$ which confirms the presence of long-range correlations in the
non-equilibrium stationary state (as already found before in the SEP \cite{S}).

The energy diffusion model is clearly very close
from the physical point of view
 to the KMP model.
But at the same time, as  we have mentioned,
it can be viewed as the continuation of the SEP
family, in the following sense:
on the one hand, the SEP can naturally  be generalized \cite{SS} to  processes
in which up to $n$ particles are allowed per site (the usual SEP has $n=1$),
and hopping rates are proportional to the number of particles at the departure
and the number of `holes' at the arrival site.
On the other hand, the energy diffusion model can be generalized to
having in each site $i$ of the lattice  $m$ free particles with momenta
$x_{i,\alpha}$, $\alpha=1,...,m$, and
the random exchange  between $x_{i,\alpha}$ and  $x_{i+1,\beta}$  leaves
$\sum_\gamma \{ x_{i,\gamma}^2 +  x_{i+1,\gamma}^2 \}$ invariant.
We shall see below that the  energy transport model with $m$ particles per
site formally corresponds to the continuation of the SEP family
to {\em negative} occupation number:  $m = -n/2$.
In other words, the energy diffusion model as defined above is, formally,
{\em `the SEP with $-1/2$ particles per site'}.

The rest of our paper is organized as follows. In section \ref{model}
we define our energy diffusion model. In section \ref{duaal} we construct
the dual process of interacting random walkers and derive the basic
corollaries of duality, namely, existence and uniqueness
of the stationary measure and expression of the stationary correlation functions
in terms of absorption probabilities. In sections 4 and 5 we derive the
exact stationary temperature profile and energy-energy correlation function.
In section 6 we show that our model satisfies local equilibrium.   
In section 7 we show a formal connection between our model and the
SEP family, showing that the energy diffusion model can be viewed as a bosonic
version of the SEP. In the last section we discuss possible asymmetric extensions
of our model.

\section{The model}\label{model}

The model is defined as a stochastic process
on $N$-dimensional vectors $(x_1,\ldots,x_N)\in\R^{N}$ which
have to be interpreted as momenta associated to lattice sites $\{1,\ldots,N\}$.
Additionally, lattice sites $1$ and $N$
are in contact with a heat reservoir at temperature $T_L$ resp.\ $T_R$.

The process is defined by its generator $L$ (acting on the core of
$\ce^\infty$ functions $f$ with compact support)

\be\label{geni}
L= L_1 + L_N + \sum_{i=1}^{N-1} L_{i,i+1}
\ee
with
\be
L_1f= T_L \frac{\partial^2 f}{\partial x_1^2} -x_1\frac{\partial f}{\partial x_1}
\ee
\be
L_N f= T_R \frac{\partial^2 f}{\partial x_N^2} -x_N\frac{\partial f}{\partial x_N}
\ee
\be
L_{i,i+1}f= \left(x_i\frac{\partial}{\partial x_{i+1}} -x_{i+1}\frac{\partial}{\partial x_{i}}\right)^2
(f)
\ee

\noindent
This corresponds, in the language of Fokker-Planck equation (or master equation) to the following
evolution equation for the time-dependent probability density $p(x,t)$:
\be\label{fok}
\frac{\partial p(x,t)}{\partial t}
= L^* p(x,t)
\ee
where $L^*$ is the adjoint (in $L^2 (dx)$) of $L$, i.e., more explicitly,
\[
L^*= L^*_1 + L^*_N + \sum_{i=1}^{N-1} L_{i,i+1}
\]
\[
L^*_1f= T_L \frac{\partial^2 f}{\partial x_1^2} + \frac{\partial}{\partial x_1}( x_1f)
\]
\[
L^*_N f= T_R \frac{\partial^2 f}{\partial x_N^2} + \frac{\partial}{\partial x_N} (x_N f)
\]

Let us first motivate the choice of this generator \eqref{geni}. The $L_1$ and $L_N$ part is the generator
of the usual
Ornstein-Uhlenbeck process which represents
thermalising noise corresponding to heat baths at temperatures $T_L$, resp.\ $T_R$.

To explain the other part, consider the operator
\be\label{boem}
{\cal A}=\left(x\frac{\partial}{\partial y}-y\frac{\partial}{\partial y}\right)^2
\ee
In polar coordinates $x=r\cos \theta$, $y=r\sin\theta$, this operator simply reads
\[
{\cal A} = \frac{\partial^2}{\partial \theta^2}
\]
which means that in the process $(x(t),y(t))$ corresponding
to \eqref{boem}, $r(t)= r(0)$ and $\theta(t)$ performs
a Brownian motion on the interval $[0,2\pi]$. More precisely, from
the It\^{o} formula, it follows that the
generator \eqref{boem} corresponds to the stochastic
differential equations (in It\^{o} sense)
\beq\label{stochboem}
dx(t) &=& -x(t) dt +\sqrt{2}\, y(t) dB(t)\nonumber\\
dy(t) &=& -y(t) dt -\sqrt{2}\, x(t) dB(t)
\eeq
where $B(t)$ is standard Brownian motion. In other words
$r^2(t)= x^2(t) + y^2(t)= r^2(0)$ with probability one
and the angular variable
$\theta(t) = \arctan (y(t)/x(t))$ is a martingale.

The bulk part $\sum_{i} L_{i,i+1}$ of the process $(x_1 (t),\ldots,x_N(t))$
corresponds then to the stochastic differential equations
\beq\label{genstochboem}
dx_1 (t)&=& -x_1(t) dt + \sqrt{2} x_2(t) d B_{1,2} (t)
\nonumber\\
dx_i(t) &=& -2x_i(t) dt +\sqrt{2}\, x_{i+1}(t) dB_{i,i+1}(t)-\sqrt{2}\, x_{i-1}(t) dB_{i-1,i} (t)
\qquad i\in [2,N-1]
\nonumber\\
dx_N (t)&=& -x_N(t) dt - \sqrt{2} x_{N-1}(t) d B_{N-1,N} (t)
\eeq
where $B_{i,i+1}(t)$ $(i=1,\ldots,N-1)$ are independent Brownian motions and $B_{0,1}(t)= B_{N,N+1}(t)=0$.
In this process, the total kinetic energy $\sum_i x_i^2(t)$ is conserved, i.e.,
$\sum_i x_i^2(t)=\sum_i x_i^2(0)$ with probability one as can be seen easily from
It\^{o}'s formula.

The full process, i.e., the process with generator \eqref{geni} (including the boundary terms $L_1, L_N$)
corresponds to the system of stochastic differential equations
\beq\label{genstochboem1}
dx_1 (t)&=& -2x_1(t) dt + \sqrt{2} x_2(t) d B_{1,2} (t) + \sqrt{2T_L} dW(t)
\nonumber\\
dx_i(t) &=& -2x_i(t) dt +\sqrt{2}\, x_{i+1}(t) dB_{i,i+1}(t)-\sqrt{2}\, x_{i-1}(t) dB_{i-1,i} (t)
\qquad i\in [2,N-1]
\nonumber\\
dx_N (t)&=& -2x_N(t) dt - \sqrt{2} x_{N-1}(t) d B_{N-1,N} (t) +\sqrt{2T_R} dW'(t)
\eeq
where $W(t),W'(t)$ are two independent Brownian motions, independent
of all the other $B_{i,i+1}(t)$.



\section{ Duality}\label{duaal}
The main tool which considerably simplifies the analysis
of this model is duality.
First note that the equations for the evolution of
correlation functions of degree $n$ for the $x$ process are {\em closed},
i.e., the time derivative of the expectation of a  polynomial
of degree $n$ in the variables $x_1,\ldots,x_n$ does not involve expectations of polynomials
of higher order \cite{GK}.

We now show that the evolution in time of well-chosen polynomials
reduces to a Markovian evolution of their {\em indices}. If we interpret
the indices of the polynomials as discrete (indistinguishable) particle configurations
(i.e., specifying for each site $i\in \{0,\ldots,N+1\}$ the number of particles), then the evolution
of the indices
turns out to become
a jump process that conserves
the total number of particles. This jump process is called the dual process.
A similar situation arises in the case of
the SEP (which is self-dual) \cite{lig}, in the case of
an infinite system of independent random walkers (where the Poisson polynomials
have the self-dual property) \cite{landim}, and in the KMP model \cite{KMP}.

We index our polynomials by a vector $\xi = (\xi_0,\ldots,\xi_{N+1})$,
$\xi_i\in\N$ and  introduce the notation
$\vec{x} = (x_0,x_1,\ldots,x_N,x_{N+1})$, where the first and last components
are {\em fixed} by
$x_0= \sqrt{T_L}$ and $x_{N+1}=\sqrt{T_R}$.
The polynomial $D(\xi,\vec{x})$ is then defined
by
\be\label{pol1}
D(\xi,\vec{x})= T_L^{\xi_0} T_R^{\xi_{N+1}}\prod_{i=1}^N \frac{x^{2\xi_i}}{(2\xi_i-1)!!}
\ee
Only even powers of $x_i$ need to be considered, since other stationary
correlations
vanish due to the invariance of the generator under the transformation
$x_i\to -x_i$.
We interpret $\xi\in\Omega=\N^{N+2}$
as prescribing the number of particles in each lattice
site $i\in\{0,\ldots,N+1\}$.

In order to introduce the generator of the dual process,
we define, for $\xi\in\Omega$, $i,j\in\{0,\ldots,N+1\}$,
the configuration $\xi^{i,j}$
to be the configuration obtained from
$\xi$ by removing one particle at $i$ and adding one
particle at $j$.
On the dual variables $\xi\in\Omega$, we then define the
following generator.
\begin{eqnarray}
\label{dualgenerator}
\loc \phi (\xi) & := &
2\xi_1 [\phi(\xi^{1,0})-\phi(\xi)]
 +  2\xi_{1}(2\xi_{2}+1) [\phi(\xi^{1,2})-\phi(\xi)]\nonumber
\\
& + & \sum_{i=2}^{N-1} \Big(2\xi_{i}(2\xi_{i-1}+1) [\phi(\xi^{i,i-1})-\phi(\xi)] + \; 2\xi_{i}(2\xi_{i+1}+1) [\phi(\xi^{i,i+1})-\phi(\xi)]\Big)\nonumber
\\
& + & 2\xi_{N}(2\xi_{N-1}+1) [\phi(\xi^{N,N-1})-\phi(\xi)]
+ 2\xi_{N} [\phi(\xi^{N,N+1})-\phi(\xi)]
\end{eqnarray}
where $\phi:\Omega\to\R$ is an arbitrary function of the finite particle configurations.

In words, we can describe the process generated by $\loc$ as follows: a configuration
$\xi=(\xi_0,\ldots,\xi_{N+1})$ represents $K$
particles (or walkers) on $\{0,1,\ldots,N+1\}$ with $K=\sum_{i=0}^{N+1} \xi_i$. The walkers
can only jump to neighboring sites and are stuck when arriving to
sites $0$ or $N+1$. The rate at which there is a jump of a walker depends on how many walkers there are
at neighboring sites.
If we have $\xi_i$ walkers at site $i$, $\xi_{i-1}$ walkers at
site $i-1$ and $\xi_{i+1}$ walkers at site $i+1$ (for $i=2,\ldots N-1$) then each of the walkers at site $i$ jumps to
site $i-1$ at rate $2(2\xi_{i-1}+1)$ and to site $i+1$ at rate $2(2\xi_{i+1}+1)$.
At the boundaries: each of the $\xi_1$ walkers at site $1$ is absorbed at site $0$ at rate $2$ and it jumps
to site $2$ at rate $2(2\xi_{2}+1)$; each of the $\xi_N$ walkers at site $N$ is absorbed at site $N+1$
at rate $2$ and it jumps to site $N-1$ at rate $2(2\xi_{N-1}+1)$.
Note that this process conserves the number of particles, i.e.,
\[
|\xi(t)| = \sum_{i=0}^{N+1} \xi_i (t) = \sum_{i=0}^{N+1} \xi_i (0) = |\xi(0)|
\]

For a single particle, i.e., $\xi = \delta_i$, the dual process is then
$\xi(t) = \delta_{X(t)}$ where $X(t)$ is a continuous time simple symmetric
nearest neighbor random walk jumping at rate 2 and absorbed upon hitting
$0$ or $N+1$. For two particles, i.e., $\xi= \delta_i + \delta_j$
the particles perform independent symmetric nearest neigbor random
walks at rate 2,
{\em except} when they are sitting at {\em neighboring sites}. In that case, i.e.,
if the two walkers are at neigboring places, one of them jumps
to the place of the other one at rate 6 (and other jumps are still at rate 2).

\br
This attractive interaction between the dual walkers has to be compared
with the repulsive (hard-core) interaction between the walkers in the
SEP.
\er

In order to formulate our duality result and its consequences, we
denote by $\hat\E_\xi$ expectation in the dual process
(i.e., the process with generator
\eqref{dualgenerator})
starting from $\xi$.
We now can formulate the duality result.
\bt
Let $\xi(t)$ denote the process with generator
\eqref{dualgenerator} and $\bix (t)$ the process
$(x_0(t),x_1(t),x_2(t),\ldots,x_N(t),x_{N+1}(t))$ where
$(x_1(t),\ldots,x_N(t))$ is the process with generator \eqref{geni},
and where $x_0(t)=\sqrt{T_L}$, $x_{N+1}(t)=\sqrt{T_R}$.
Then we have
\be\label{dualityrel}
\E_{\bix} [D(\xi,\bix(t))] = \hat\E_\xi [D(\xi(t),\bix)]
\ee
\et
\vspace{.4cm}
\bpr
Start from \eqref{geni}.
For $i=1,\ldots,N-1$ we have
\begin{eqnarray*}
&&
L_{i,i+1} D(\xi,x)
=
T_L^{\xi_0}\;T_R^{\xi_{N+1}} \left(\prod_{k\not\in\{i,i+1\}}  \frac{x_k^{2\xi_k}}{(2\xi_k -1)!!}\right)
\times
\\
&&\left(2\xi_{i+1}(2\xi_{i+1}-1) \frac{x_i^{2\xi_i+2}}{(2\xi_i -1)!!}\frac{x_{i+1}^{2\xi_{i+1}-2}}{(2\xi_{i+1} -1)!!}
- 2\xi_{i}(2\xi_{i+1}+1) \frac{x_i^{2\xi_i}}{(2\xi_i -1)!!}\frac{x_{i+1}^{2\xi_{i+1}}}{(2\xi_{i+1} -1)!!}
\right.
\\
&&\left.- 2\xi_{i+1}(2\xi_{i}+1) \frac{x_i^{2\xi_i}}{(2\xi_i -1)!!}\frac{x_{i+1}^{2\xi_{i+1}}}{(2\xi_{i+1} -1)!!}
+2\xi_{i}(2\xi_{i}-1) \frac{x_i^{2\xi_i-2}}{(2\xi_i -1)!!}\frac{x_{i+1}^{2\xi_{i+1}+2}}{(2\xi_{i+1} -1)!!}
\right)
\\
\end{eqnarray*}
which implies
\begin{eqnarray*}
L_{i,i+1} D(\xi,x)
& = &
\Big(2\xi_{i+1}(2\xi_{i}+1) [D(\xi^{i+1,i},x)-D(\xi,x)]
\\
&&
\;+\;2\xi_{i}(2\xi_{i+1}+1) [D(\xi^{i,i+1},x)-D(\xi,x)]\Big)
\\
\end{eqnarray*}
Furthermore
\begin{eqnarray*}
L_1 D(\xi,x) & = & T_R^{\xi_{N+1}} \left(\prod_{k\not\in\{1\}}  \frac{x_k^{2\xi_k}}{(2\xi_k -1)!!}\right) \times
\\
&&
\left(
T_L^{\xi_0 +1}\;2\xi_1 (2\xi_1 -1)\frac{x_{1}^{2\xi_{1}-2}}{(2\xi_{1} -1)!!}
- T_L^{\xi_0}\;2\xi_1 \frac{x_{1}^{2\xi_{1}}}{(2\xi_{1} -1)!!}
\right) \nonumber\\
&=&
2\xi_1 [D(\xi^{1,0},x)-D(\xi,x)]
\end{eqnarray*}
and
\begin{eqnarray*}
L_N D(\xi,x) & = & T_L^{\xi_0} \left(\prod_{k\not\in\{N\}}  \frac{x_k^{2\xi_k}}{(2\xi_k -1)!!} \right)\times
\\
&&
\left(
T_R^{\xi_{N+1}+1} \;2\xi_N (2\xi_N -1)\frac{x_{N}^{2\xi_{N}-2}}{(2\xi_{N} -1)!!}
- T_R^{\xi_{N+1}} \;2\xi_N \frac{x_{N}^{2\xi_{N}}}{(2\xi_{N} -1)!!}
\right)\nonumber\\
&=&
2\xi_{N} [D(\xi^{N,N+1},x)-D(\xi,x)]
\end{eqnarray*}

Therefore, we obtain
\be\label{dualcom}
L D(\xi,\bix) = \loc D(\xi,\bix)
\ee
where in the lhs of \eqref{dualcom} the operator
$L$ is working on the $\bix$ variable, and
in the rhs the operator $\loc$ is working on
the $\xi$ variable.

This relation then lifts to the semigroups and the processes via a standard
argument, see e.g.\ the proof of theorem 1.1 p.\ 363 in
\cite{lig}. (Notice that informally, by \eqref{dualcom}
$\E_x D(\xi,\bix(t))= e^{tL} D(\xi,\cdot) (\bix)= e^{t\loc} D(\cdot,\bix) (\xi)
= \hat\E_\xi D(\xi(t),\bix)$).
\epr
Notice that in the dual process all particles are eventually
absorbed at one of the boundaries $0, N+1$. Hence, the limiting
configuration of the dual process starting from $\xi\in\Omega$
will always be of the form
$k\delta_0 + l \delta_{N+1}$ where $k+l= \sum_{i} \xi_i$.
We say that a function $f: \Omega\to\R$ is harmonic
for the process generated by $\loc$ if $f(\xi) = \hat\E_\xi [f(\xi(t))]$.
Since the configurations of the type $k\delta_0 + l \delta_{N+1}$ are
absorbing, the corresponding absorption probabilities
\be\label{absoco}
c_{kl} (\xi) =\hat\pee_\xi (\xi(\infty)=k\delta_0 + l \delta_{N+1})
\ee
are harmonic.

The following propositions gives some
consequences of duality.
\bp If $T_L=T_R=T$, then the unique
stationary measure of the process
$x(t) = (x_1(t), \ldots, x_N(t))$ is the Gaussian product measure with probability density function
\be\label{gausseq}
\rho_T(x) = \frac{1}{(2\pi T)^{N/2}}\exp\left(-\sum_{i=1}^N \frac{x_i^2}{2T}\right)
\ee
From every initial condition $(x_1,\ldots,x_N)$ the process
converges in distribution to $\rho_T$.
\ep
\bpr
We first show stationarity of $\rho_T$.
If $T_L=T_R=T$, for the Gaussian measure \eqref{gausseq}
we have
\[
\int D(\xi, \bix) \rho_T(x) dx = T^{|\xi|}
\]
where $|\xi|= \sum_{i=1}^N \xi_i$.
Therefore, using \eqref{dualityrel}
\beq\label{bom}
\int \E_{\bix}[D(\xi, \bix(t))] \,\rho_T (x) dx
&=&\hat\E_\xi \int D(\xi(t), \bix) \rho_T(x) dx
\nonumber\\
&=& \hat \E_\xi T^{|\xi(t)|}
\nonumber\\
&=& T^{|\xi|}
=\int D(\xi, \bix) \rho_T(x) dx
\eeq
where in the third equality we used that in the dual process the
number of particles is conserved.
Next, to prove the convergence, remember that in the dual
process eventually all particles are absorbed at $0$ or at $N+1$. Therefore,
if $T_L=T_R=T$,
\beq\label{bombom}
\lim_{t\to\infty}\E_{\vec{x}} (D(\xi, \bix(t)) &=& \sum_{k,l: k+l= |\xi|}
\hat\pee_\xi (\xi(\infty) = k\delta_0 + l\delta_{N+1})T^k T^l=
T^{|\xi|}
\nonumber\\
&=&
\int D(\xi, \bix) \rho_T(x) dx
\eeq
\epr
\bp
For all $T_L \neq T_R$, there is a unique stationary measure $\mu_{T_L, T_R}$
which has finite moments of every order, and for this measure one has
\be\label{abso}
\muT ( D(\xi,\bix)) = \sum_{k,l\;: \,k+l= |\xi|} T_L^k \,T_R^l\, c_{kl} (\xi)
\ee
where $c_{kl}(\xi)$ are the absorption probabilities defined in \eqref{absoco}.
\ep
\bpr
If $\muT$ is invariant and has finite
moment of all order, then, by \eqref{dualityrel}
we have
\beq\label{bombombom}
\int D(\xi,\bix) \muT (dx)&=& \lim_{t\to\infty} \int\E_x(D(\xi,\bix(t)))\muT(dx)
\nonumber\\
&=&
\int \left(\lim_{t\to\infty}\hat{\E_\xi} (D(\xi(t),\bix))\right)\muT(dx)
\nonumber\\
&=&
\sum_{k,l\;: \,k+l= |\xi|} T_L^k T_R^l\hat\pee_\xi (\xi(\infty) = k\delta_0 + l\delta_{N+1})
\eeq
where in the last equation we used that eventually all particle are absorbed at one
of the absorbing states $0$ or $N+1$.
Conversely, if \eqref{abso} holds, then, since the
absorption probabilities are harmonic for the dual process, by \eqref{dualityrel}
we have
\[
\int D(\xi,\bix) \mu_{T_L,T_R} (dx)=\int \hat\E_\xi [D(\xi(t), \vec{x})] \mu_{T_L,T_R}(dx)
=\int \E_{\bix} [D(\xi, \vec{x}(t))] \mu_{T_L,T_R}(dx)
\]
which shows invariance of $\muT$.
\epr

\section{Temperature profile.}\label{tf}
The following proposition shows the convergence to
and linearity of the limiting  temperature profile
(linearity of the stationary temperature profile has been shown
before in \cite{GK}).
\bp
\noindent
\begin{itemize}
\item[a)] Starting from $x\in\R^N$, the local temperature
at site $i$ satisfies
\be\label{randwalk}
\E_x (x_i^2 (t))= \hat\E_i (x_{X(t)}^2)= \sum_{j} p_t(i,j) x_j^2
\ee
where $X(t)$ is
continuous-time
simple symmetric random walk jumping at rate 2, and absorbed at $0$, $N+1$,
and where $p_t(i,j)$ denotes the transition probability
of this random walk to go from $i$ to $j$ in
time $t$.
\item[b)] The stationary temperature profile is given
by
\beq\label{limittemp}
T_i:= \muT (x_i^2)  &=&
T_L \,\hat\pee_i(X(\infty) =0) + T_R\, \hat\pee_i(X(\infty) =N+1)
\nonumber\\
&=&
T_L \left(1-\frac{i}{N+1}\right) +T_R \left(\frac{i}{N+1}\right)
\eeq

\item[c)] For the stationary local temperature autocorrelation
function we have
\be\label{twopoint}
\E_{\muT} (x^2_i(0) x^2_j(t)) -T_i T_j=
\sum_{k=1}^N \hat\pee_j (X_t=k, \tau>t) \muT (x_i^2 x_k^2)
- T_iT_j \hat\pee_j (\tau >t)
\ee
where $\tau$ is the absorption time of the random walk
$\{ X_t:t\geq 0\}$. As a consequence,
\be\label{quasi}
\lim_{t\to\infty} \frac{\E_{\muT} (x^2_i(0) x^2_j(t)) -T_i T_j}{\hat\pee_j (\tau>t)}
= -T_iT_j + \frac1N \sum_{k=1}^N \muT (x_i^2 x_k^2)
\ee
\end{itemize}
\ep
\bpr

\noindent
\begin{itemize}
\item[a)] This follows from \eqref{dualityrel} and
the fact that the dual process starting from a single
particle at $i$ is given by the random walk $X(t)$.
\item[b)] This follows by taking the limit $t\to\infty$ in
\eqref{randwalk} and the well-known formula for the absorption probabilities
of the random walk $X(t)$.
\item[c)] The first statement \eqref{twopoint} follows from
\eqref{randwalk} and \eqref{limittemp}. The second statement
\eqref{quasi} follows from the fact that the quasistationary distribution
of the random walk $X(t)$ is the uniform measure on $\{ 1,\ldots,N\}$, i.e.,
for any function $f: \{1,\ldots,N\} \to \R$:
\[
\lim_{t\to\infty} \hat\E_i (f(X_t)|\tau>t) = \frac1N \sum_{j=1}^N f(j)
\]
\end{itemize}
\epr

\noindent

\section{Stationary two-point correlation function}
We now study the two point function $\muT(x_i^2 x_j^2)$. In the dual process we then have
two particles starting at sites $i$ and $j$ of which we denote the positions at time $t$
by $(X_t,Y_t)$. From (\ref{dualgenerator}) we see that these two particles behave
as two independent random walkers (with rate 2) as long as they are not at neighboring
sites. When they are at neighbor sites then on the next step they prefer to
be on same site compared to being further separated.

\noindent
From \eqref{abso} we infer
\beq\label{twoabso}
\muT (x_i^2 x_j^2) &=& (1+2\delta_{i,j})\Big( T_L^2 \,
\hat\pee_{ij}( \{X_\infty,Y_\infty\}= \{0\})
+ T_R^2\, \hat \pee_{ij}( \{X_\infty,Y_\infty\}= \{N+1\})
\nonumber\\
&+& T_L T_R \,\hat \pee_{ij} (\{X_\infty,Y_\infty\}= \{0,N+1\})\Big)
\eeq

\noindent
Unfortunately, now the absorption probabilities cannot be obtained by elementary probabilistic
considerations since the two random walkers of the dual process are interacting
(attracting each other when they are at nearest neighbor sites). We therefore proceed by directly
solving the linear equations for
$y_{i,j} = \muT(x_i^2 x_j^2)$. We restrict ourself to the case
$1\leq i \leq j \leq N$ since $\muT (x_i^2 x_j^2)$ is obviously symmetric.
We further denote $y_i=T_i= \muT(x_i^2)$.

The $y_{i,j}$ then satisfy the following system of linear equations:
\begin{eqnarray}
& & 3\,y_{i - 1, i} -2\, y_{i,i} + 3 \,y_{i, i + 1} = 0  \qquad\qquad\qquad\qquad\qquad j=i \nonumber \\
& & y_{i - 1, i + 1} +  y_{i + 1, i + 1} + y_{i,i} + y_{i, i + 2} - 8\,y_{i, i + 1} = 0  \qquad\;\; j = i+1 \nonumber \\
& & y_{i - 1, j} + y_{i + 1, j} + y_{i, j - 1} + y_{i, j + 1} - 4\, y_{i, j} = 0 \qquad\qquad j\geq i+2
\end{eqnarray}
with boundary conditions given by
\begin{eqnarray}
& & y_{i,N+1} = y_i y_{N+1} = y_i T_R \qquad\qquad i=0,\ldots,N+1
\nonumber\\
& & y_{0,j} = y_0 y_{j} = T_L y_j \qquad\qquad\qquad\, j=0,\ldots,N+1
\end{eqnarray}
Plugging in the ansatz
\[
y_{i,j} = A + Bi + Cj + Dij
\]
we find that
these equations have the following solution: if $j > i $ then
\be
y_{i,j}= T_L^2 \;+\;
i\; \frac{T_R-T_L}{N+1}\left(T_L + 2\frac{T_R-T_L}{N+3}\right) \;+\;
j\; \frac{ T_L \left(T_R -T_L\right)}{N+1} \;+\;
ij \; \frac{\left(T_R-T_L\right)^2}{(N+1)(N+3)}
\ee
and for the diagonal $i=j$ we have
\beq
y_{i,i}&= & 3\left(T_L^2-\frac{(T_R-T_L)^2}{(N+1)(N+3)}\right)
\nonumber\\
&\;+\; &
6\,i\; \frac{T_R-T_L}{N+1}\left(T_L +\frac{T_R-T_L}{N+3}\right)
\nonumber\\
 &\;+\; &
3\,i^2\;\frac{(T_R-T_L)^2}{(N+1)(N+3)}
\eeq

\noindent
The solution can also be written as
a quadratic form in $T_L$ and $T_R$ and then the coefficients of the quadratic
form add up to $1$ outside of diagonal and to $3$ on the diagonal
(as can be seen directly from \eqref{twoabso}). Explicitly
\beq
y_{i,j} & = &
\left(1-\frac{i}{N+3}\right) \left(1-\frac{j}{N+1}\right) \;T_L^2 \; + \;
\frac{i (2+j)}{(N+1)(N+3)} \;T_R^2\;+\; \nonumber \\
& & \Big[1-\left(1-\frac{i}{N+3}\right) \left(1-\frac{j}{N+1}\right)-\frac{i (2+j)}{(N+1)(N+3)}\Big]\;T_L T_R
\eeq
and
\beq
y_{i,i} & = & 3\left[\left(1-\frac{i}{N+3}\right) \left(1-\frac{i}{N+1}\right)-\frac{1}{(N+1)(N+3)}\right] \;T_L^2
\nonumber\\
&\;+\; &
3\left[\frac{i (2+i)}{(N+1)(N+3)} - \frac{1}{(N+1)(N+3)}\right]\;T_R^2
\nonumber\\
&\;+\; &
3\left[1-\left(1-\frac{i}{N+3}\right) \left(1-\frac{i}{N+1}\right)-\frac{i (2+i)-2}{(N+1)(N+3)}\right]\;T_L T_R
\eeq

Comparing this with \eqref{twoabso}
we can {\em read of} the expressions for the absorption probabilities.
E.g., for $i < j$
\be\label{stickabso}
\hat\pee_{ij} (X_\infty=0, Y_\infty=0)
=\left(1-\frac{i}{N+3}\right) \left(1-\frac{j}{N+1}\right)
\ee
Remark that this probability is larger than in the case
of independent random walkers where the expression would be
$(1-\frac{i}{N+1})(1-\frac{j}{N+1})$. This means that due to the attractive
interaction of the walkers, they have a (small) preference
to be absorbed at the same site. This effect is however
negligible (of order $1/N$) as $N\to\infty$.

\noindent
Finally, we compute the covariance
$$
c_{i,j} = \muT(x_i^2 x_j^2) - \muT(x_i^2)\muT(x_j^2)
$$
and we obtain
\be
c_{i,j}=\frac{2i(N+1-j)}{(N+3)(N+1)^2}(T_R-T_L)^2
\ee
\beq
c_{i,i}&=& \frac{(2N^2+8N+3) T_L^2+6 T_L T_R -3 T_R^2}{(N+1) (N+3)}
\;+\; i\;\frac{2 \left(T_R-T_L\right) \left((3+2 N) T_L +3 T_R\right)}{(N+1)(N+3)}
\nonumber\\
&\; + \; &i^2\;\frac{2  N \left(T_R-T_L\right)^2}{(N+1)^2 (N+3)}
\eeq
\br
Notice that the covariance $c_{i,j}$ manifests the presence of long range
correlations which are believed to be typical for non-equilibrium steady states
\cite{S}.
The fact that the covariance is positive is due to the attractive character of the interaction
between the dual walkers.
This has to be contrasted with the case of the SEP model where
the covariance is negative and the interaction between walkers is repulsive.
\er

We notice that in $c_{i,j}$ appears the Green's function of
the simple (continuous time) random walk $X_t$,
\[
 G(i,j)=\frac{2i(N+1-j)}{(N+1)}
\]
 i.e., the  expected total time spent at $i$ starting at $j$
before being absorbed at $0$ or $N+1$. Moreover, for $0<\alpha_1 <\alpha_2<1$,
\be\label{covaa}
\lim_{N\to\infty} Nc_{N\alpha_1,N\alpha_2} = 2(T_R-T_L)^2 \alpha_1 (1-\alpha_2)
\ee
From \eqref{covaa}, in the spirit of \cite{lao},
one expects that in the limit $N\to\infty$ the random distributions
\[
Y^N_\alpha:= \frac{1}{\sqrt{N}}\sum_{i=1}^N (x_i^2- T_\alpha)\delta(\alpha -i/N)
\]
converge jointly to a Gaussian random field with
covariance
\[
 C(\alpha_1,\alpha_2)= 2(T_L-T_R)^2 \alpha_1(1-\alpha_2) + 2T^2({\alpha_1}) \delta(\alpha_1-\alpha_2)
\]
where
\be\label{tempa}
T(\alpha) = \lim_{N\to\infty} T_{\lfloor\alpha N\rfloor} = (1-\alpha)T_L +  \alpha T_R
\ee
with $\lfloor\alpha N\rfloor$ denoting the integer part of $\alpha N$.

\section{Local equilibrium}
From the computation of the stationary two point correlation,
we infer in particular that for $\alpha\in (0,1)$
\[
\lim_{N\to\infty} \muT ( x^4_{\lfloor\alpha N\rfloor})
=3 T^2(\alpha)
= \frac{1}{\sqrt{2\pi T(\alpha)}}\int x^4e^{-x^2/2T(\alpha)}dx
\]
while for each $i\in\N, i>0$
\[
\lim_{N\to\infty} \muT ( x^2_{\lfloor\alpha N\rfloor} x^2_{\lfloor\alpha N\rfloor+i})
=T^2(\alpha)
=\left(\frac{1}{\sqrt{2\pi T(\alpha)}}
\int x^2e^{-x^2/2T(\alpha)}dx\right)^2
\]
where $T_\alpha$ is defined in \eqref{tempa}.

This suggests that ``around each macroscopic point $\alpha$",
which we associate to the micro-point $\lfloor\alpha N\rfloor$, there
is a Gaussian ``local equilibrium" distribution
with variance $T(\alpha)$.
More precisely we give the following definition cf. \cite{KMP}.
\bd
Let $T_L$ and $T_R$ be {\em fixed} and let $\alpha\in (0,1)$.
We say that local equilibrium holds if for
all $n\in\N$, for all $k_1,\ldots,k_n\in \N$ and
for all $i_1,\ldots, i_n\in \{1,\ldots,N\}$ {\em fixed},
we have:
\[
\lim_{N\to\infty} \int D\left(\sum_{l=1}^n k_l\delta_{\lfloor\alpha N\rfloor +i_l}, \bix\right)  \muT(dx)
=\prod_{l=1}^n \int \frac{x^{2k_l}}{(2k_l -1)!!}
\,\rho_{T(\alpha)}(dx)
= [T(\alpha)]^{\sum_{l=1}^n k_l}
\]
where $\rho_{T(\alpha)}$ denotes the measure of a centered guassian
variable with variance $T(\alpha)$.
\ed
This is equivalent with the requirement that
\be
\lim_{N\to\infty} \tau_{\lfloor\alpha N\rfloor}(\muT)
= \caG(T(\alpha))
\ee
where $\tau_{\lfloor\alpha N\rfloor}$ denotes spatial
shift, $\caG(\sigma)$ denotes the product measure on
$\R^{\N}$ with marginals that are normally distributed
with mean zero and variance $\si^2$, and where the limit
is in the sense that expectations of
polynomials of type $D(\xi,\bix)$ converge to
the corresponding expectations in the Gaussian measure.

The following lemma shows that factorization of
the absorption probabilities is sufficient for local
equilibrium.
\bl
Let $T_L$ and $T_R$ be {\em fixed} and let $\alpha\in (0,1)$.
Suppose that for all $n\in\N$, for all $k_1,\ldots,k_n\in\N$, for all
$i_1,\ldots,i_n\in \{1,\ldots,N\}$ {\em fixed} 
and for all $K,L\in\N$ with $K+L= \sum_{l=1}^n k_l$
\begin{eqnarray}
\label{facto}
\lim_{N\to\infty} \hat\pee_{\sum_{l=1}^n k_l\delta_{\lfloor\alpha N\rfloor+i_l}}
(\xi({\infty}) = K\delta_0 + L\delta_{N+1}) \nonumber \\
= 
\lim_{N\to\infty}\hat\pee_{\lfloor\alpha N\rfloor} (X(\infty)=0)^K
\hat\pee_{\lfloor\alpha N\rfloor} (X(\infty)=N+1)^L
\end{eqnarray}
then local equilibrium holds.
\el
\bpr
Combination of \eqref{abso} with \eqref{facto} gives
\beq
&&\lim_{N\to\infty} \int D\left(\sum_{l=1}^n k_l\delta_{\lfloor\alpha N\rfloor+i_l},
\bix\right)\muT (dx)
\nonumber\\
 &=&\sum_{K,M:K+M=k_1+\ldots + k_n}
\lim_{N\to\infty}\hat\pee_{\lfloor\alpha N\rfloor} (X(\infty)=0)^M
\hat\pee_{\lfloor\alpha N\rfloor} (X(\infty)=N+1)^K T_L^M T_R^K
\nonumber\\
&=&
\lim_{N\to\infty}\left(
\hat\pee_{\lfloor\alpha N\rfloor} (X(\infty)=0) T_L
+ \hat\pee_{\lfloor\alpha N\rfloor} (X(\infty)=N+1) T_R\right)^{k_1+\ldots+ k_n}
\nonumber\\
&=&
\lim_{N\to\infty} T_{\lfloor\alpha N\rfloor}^{k_1+\ldots+ k_n}
=
T(\alpha)^{k_1+\ldots+ k_n}
\eeq
\epr
In order to see heuristically why the factorization property for the absorption probabilities holds, it
suffices to see that if we start $n$ dual random walkers
$X_1(t),\ldots, X_n(t)$
from initial positions $\lfloor \alpha N\rfloor + i_l$, $l=1,\ldots,n$
then we can couple them with $n$ independent random walkers
$X'_1(t),\ldots,X'_n(t)$
started at the same initial positions
such that for all $\epsi>0$ and for all $l\in\{ 1,\ldots,n\}$
\be\label{coup}
X_l(t)-X'_l(t)\leq \epsi \sqrt{t}
\ee
with probability close to one for $t$ large enough.
Indeed, if dual walker $l$ is absorbed at $0$,
then his happens at a time of the order
$N^2$, and then, for large $N$, \eqref{coup} tells that the corresponding independent
walker is at distance less than $\epsi N$ from $0$ at that time.
Therefore, the probability that this independent walker is absorbed at $N+1$
is at most $\epsi N/(N+1)\leq \epsilon$. So the probability (in the coupling)
that dual walker $X_l$ and corresponding independent walker $X'_l$ are
absorbed at different points is less than $\epsi$. Therefore, if we have the coupling
with property \eqref{coup}, we have for all $\theta_1,\ldots,\theta_n\in \{0,N+1\}$
\[
 \lim_{N\to\infty} \Big(\hat\pee
(X_1(\infty)=\theta_1,\ldots,X_n(\infty)=\theta_l)
-
\pee'
(X'_1(\infty)=\theta_1,\ldots,X'_n(\infty)=\theta_l)\Big)=0
\]
where $\hat\pee$ refers to the probability
measure on path space for dual walkers starting at positions
$\lfloor\alpha N\rfloor +i_1,\ldots,\lfloor\alpha N\rfloor +i_n$, and
$\pee'$
refers to the probability for independent walkers starting at positions
$\lfloor\alpha N\rfloor +i_1,\ldots,\lfloor\alpha N\rfloor +i_n$.

To see that such a coupling exists, we observe that in the dual process
there is only interaction of the walkers when they are at neigboring positions.
In that case they jump with higher rate (than independent walkers)
to the same position. The coupling then consists in letting
the independent walkers and the dual walkers perform the same jumps,
and having extra jumps for the dual walkers
(which are not performed by the independent walkers) when they are at neigboring
positions. The total time that dual walkers are at neigboring
positions in the time interval $[0,t]$ is less
than $t^{\frac12 +\delta}$ with probability close to one. So the difference
between the position of the independent walker and the dual walker
is a sum of the order of $t^{\frac12 +\delta}$  independent increments
taking values $\pm 1$,
which is bounded by $(t^{\frac12 +\delta})^{\frac12 +\delta}\leq t^{\frac12 -\delta'}$ with high probability.

A similar idea of coupling has been implemented in the context of
the simple symmetric exclusion process,
see \cite{pres}.
For the full proof of the factorization property along these lines we however
refer to future work.

\section{The energy-diffusion model and the SEP family: bosons versus fermions.}

In this section we shall see the formal relation between the SEP family
and the energy-diffusion model. This can help as a guide to
see the similarities in methods for treating both cases.

We shall first consider a generalization of the energy-diffusion model
to the case in which  there are $m$ momenta per site,
and kinetic energy is exchanged between any two momenta in neighboring sites.
The generator is again (\ref{geni}), with now
\be
L_1= \sum_{\alpha=1}^{m} \left [ T_L
\frac{\partial^2}{\partial x_{1,\alpha}^2} -
 x_{1,\alpha} \frac{\partial}{\partial x_{1,\alpha}}
\right ]
\ee
\be
L_N=  \sum_{\alpha=1}^m \left[
T_R \frac{\partial^2}{\partial x_{N,\alpha}^2} -
x_{N,\alpha} \frac{\partial}{\partial x_{N,\alpha}}\right]
\ee
\be
L_{i,i+1}= \frac{1}{m} \sum_{\alpha,\beta=1}^m
\left(x_{i,\alpha}
\frac{\partial}{\partial x_{i+1,\beta}} -
x_{i+1,\beta}\frac{\partial}{\partial x_{i,\alpha}}
  \right)^2
\label{bb}
\ee
The factor $1/m$ multiplying $L_{i,i+1}$ is rather arbitrary,
it sets the time-scale.
We wish to show the connection of this family of models (labelled by $m$)
with the (partial) exclusion process, with maximal occupancy $n$,
and in which the jumping rate
is proportional both to the number of particles on the
departure configuration, and to the number of `holes' ($n$ minus
the number of particles)
in the arrival configuration.
The evolution operator of this process can be written as the Hamiltonian
$H$ of the spin $j=n/2$ ferromagnet \cite{SS}
\be
H=-L_{SEP}^*
\ee
with
\begin{eqnarray}
L^*_{SEP} &=& \frac{1}{j}
  \sum_i  \left(J^+_i J^-_{i+1} + J^-_i J^+_{i+1} + 2 J^o_i J^o_{i+1}
 - 2 j^2  \right)\\
&+&\alpha (J^-_1 - J^o_1-j) + \gamma (J^+_1 + J^o_1-j)
+ \delta (J^-_L - J^o_L-j) + \beta (J^+_L + J^o_L-j)\nonumber
\end{eqnarray}
The factor $1/j$ is analogous to the factor $1/m$ in (\ref{bb}).
The operators $J^+_i, J^-_i, J^o_i$ act on the Hilbert space
 corresponding to  $0 \le r \le n$  particles per site $\otimes_i |r\rangle_i$
as follows:
\begin{eqnarray}
J^+_i |r\rangle_i &=& (2j-r)  |r\rangle_i \nonumber \\
 J^-_i |r\rangle_i &=& r  |r\rangle_i \nonumber \\
J^o_i |r\rangle_i &=& (r-j)  |r\rangle_i
\end{eqnarray}
They satisfy the commutation relations of the $SU(2)$ algebra:
\begin{equation}
[J_i^o,J_i^\pm]=\pm J_i^\pm \;\;\;\;\; ; \;\;\;\;\;  [J_i^-,J_i^+]=-2J_i^o
\end{equation}
and they can be transformed to the conventional $SU(2)$ matrices by a similarity
transformation.

Representations are labeled by the squared angular momentum operator:
 \begin{equation}
J^2 |jM> = j(j+1)  |jM>
\end{equation}
with $j=n/2$, so that
the ordinary SEP
(with $(0,1)$ occupation) corresponds to
a representation of spin $1/2$.

Going back to the generalized model defined above, we can rewrite the generator
of the energy diffusion process as the Hamiltonian $H=-L^*$ with
\begin{eqnarray}
L^*&=& \frac{4}{m} \sum_i \left(
K^+_i K^-_{i+1} + K^-_i K^+_{i+1} - 2 K^o_i K^o_{i+1}
+ \frac{m^2}{8} \right)
\nonumber\\
&+&2 \left(T_1 K^+_1 - K^o_1 - \frac{m}{4}\right)
+2 \left(T_L K^+_L - K^o_L -\frac{m}{4}\right)
\end{eqnarray}
where we have defined, in each site:
\begin{eqnarray}\label{Koper}
K^+_i &=& \frac{1}{2} \sum_\alpha x_{i,\alpha}^2 \nonumber \\
K^-_i &=& \frac{1}{2} \sum_\alpha
\frac{\partial^2}{\partial x_{i,\alpha}^2} \nonumber \\
K^o_i&=& \frac{1}{4} \sum_\alpha \left\{
\frac{\partial}{\partial x_{i,\alpha}} x_{i,\alpha} +
 x_{i,\alpha} \frac{\partial}{\partial x_{i,\alpha}} \right \}
\end{eqnarray}
These satisfy the SU(1,1) relations:
\begin{equation}
[K_i^o,K_i^\pm]=\pm K_i^\pm \;\;\;\;\; ; \;\;\;\;\;  [K_i^-,K_i^+]=2K_i^o
\end{equation}
(note the sign difference with respect to $SU(2)$).
Representations are labeled in a manner analogous to $SU(2)$:
\begin{eqnarray}
K_i^2 |kM> &=&
 [K_i^o]^2 - \frac{1}{2} [K_i^+K_i^-+K_i^-K_i^+]|kM>= k(k-1)|kM>\nonumber \\
 K_i^o|kM>&=&m|kM>
\end{eqnarray}
To identify which is the representation (i.e., the value of $k$),
we check the value of $K_i^2$ as applied to the constant (which is the zero
eigenvalue eigenvector of $L^*$):
\begin{equation}
K_i^- |1>=0 \;\;\; ; \;\;\; K_i^o |1>=k|1>=\frac{m}{4}|1>  \;\;\; ; \;\;\;
K_i^2 |1>=k(k-1)|1>=\frac{m}{4}(\frac{m}{4}-1)|1>
\end{equation}
Hence $k=m/4$, and in particular $k=1/4$ for the process with one velocity
 per site.

 Consider the coherent-state representation of vectors and operators (cf.\cite{coherent, coherent1,coherent2}):
\begin{eqnarray}
 \psi(z_i) &=& <z_i|\psi> \;\;\; \mbox{with} \;\;\; |z_i>=e^{z_i^* K_i^+}|0>
\nonumber \\
 \psi(z_i) &=& <z_i|\psi> \;\;\; \mbox{with} \;\;\; |z_i>=e^{z_i^* J_i^+}|0>
\end{eqnarray}
(where $|0>$ is the state anihilated by $J^-_i$ or $K^-_i$).
The group operators act on such states as:
\begin{eqnarray}
<z_i|K^+_i|\psi> &=& \left[z_i^2 \frac{\partial}{\partial z_i}+ 2kz_i
\right] <z_i|\psi> \nonumber \\
<z_i|K^-_i|\psi> &=&  \frac{\partial}{\partial z_i} <z_i|\psi> \nonumber \\
 <z_i|K^o_i|\psi> &=& \left[z_i \frac{\partial}{\partial z_i}+ k
\right] <z_i|\psi>
\end{eqnarray}
and
\begin{eqnarray}
<z_i|J^+_i|\psi> &=& \left[-z_i^2 \frac{\partial}{\partial z_i}+ 2jz_i
\right] <z_i|\psi> \nonumber \\
<z_i|J^-_i|\psi> &=&  \frac{\partial}{\partial z_i} <z_i|\psi> \nonumber \\
 <z_i|J^o_i|\psi> &=& \left[z_i \frac{\partial}{\partial z_i}-j
\right] <z_i|\psi>
\end{eqnarray}

Now, writing the generator of the  energy diffusion model and the SEP models
in these representations, it is easy to check that models with
$k=m/4$
of one class  formally map into models with $-j=-n/2$
 of the other class, albeit with   somewhat different
boundary terms.

Another way to see the relation between these two families of models
 is to note that we can rewrite the generator $L$
of the energy transport model {\em with even } $m$ in terms of
$m/2$ bosons $a_{i\alpha}$ and $m/2$ bosons $b_{i\alpha}$ in each site
\begin{equation}
 a_{i\alpha} a^\dag_{j\beta} - a^\dag_{j\beta} a_{i\alpha} = \delta_{ij}
\delta_{\alpha \beta}\;\;\; ; \;\;\;
 b_{i\alpha} b^\dag_{j\beta} - b^\dag_{j\beta} b_{i\alpha} = \delta_{ij}
\delta_{\alpha \beta}
\end{equation}
(all other commutators vanish) and writing
\begin{eqnarray}
K^+_i &=& \sum_\alpha a^\dag_{i\alpha} b^\dag_{i\alpha}
 \;\;\; ; \;\;\; K^-_i = \sum_\alpha b_{i\alpha} a_{i\alpha} \nonumber \\
K^o_i &=& \frac{1}{2} \sum_\alpha
\left( a^\dag_{i\alpha} a_{i\alpha} +  b^\dag_{i\alpha} b_{i\alpha} \right)
+r
\end{eqnarray}
We can similarly write the generalized SEP with allowed occupancy $n$
with
$n$ {\em fermions}
 $a_{i\alpha}$ and $n$ fermions $b_{i\alpha}$ in each site
\begin{equation}
 a_{i\alpha} a^\dag_{j\beta} + a^\dag_{j\beta} a_{i\alpha} = \delta_{ij}
\delta_{\alpha \beta}\;\;\; ; \;\;\;
 b_{i\alpha} b^\dag_{j\beta} + b^\dag_{j\beta} b_{i\alpha} = \delta_{ij}
\delta_{\alpha \beta}
\end{equation}
(all other anti-commutators vanish) and writing
 \begin{eqnarray}
J^+_i &=& \sum_\alpha a^\dag_{i\alpha} b^\dag_{i\alpha}
 \;\;\; ; \;\;\; J^-_i = \sum_\alpha b_{i\alpha} a_{i\alpha} \nonumber \\
J^o_i &=& \frac{1}{2} \sum_\alpha
\left( a^\dag_{i\alpha} a_{i\alpha} +  b^\dag_{i\alpha} b_{i\alpha} \right)
-n
\end{eqnarray}

Hence, SEP and energy diffusion models are essentially
fermionic and bosonic counterparts
of the same models. This is perhaps not surprising,
in view of the fact that Poisson processes (as the SEP)
are related to Fermions, just as Gaussian processes are related
to bosons \cite{BF, BF1}.

\section{Asymmetric generalizations}

In this last section we briefly indicate how to introduce asymmetry in our model
without breaking the bulk energy conservation law.
This is in the spirit of introducing a bias to move in a preferred
direction in the simple symmetric exclusion process.

Consider as the first example two neighboring sites with momenta $x,y$.
Going back to Eq. (\ref{boem}), we may add to the diffusion term a bias:
\be
{\cal A}^{drift}=
\left(x\frac{\partial}{\partial y}-y\frac{\partial}{\partial x}\right)^2
+  E(x,y)
\left(x\frac{\partial}{\partial y}-y\frac{\partial}{\partial x}\right)
\ee
and its obvious generalization to a chain then becomes the asymmetric analog
of the bulk part of \eqref{geni}
\be\label{obviousge}
\sum_{i}L^{drift}_{i,i+1}=
\sum_i\left(x_{i}\frac{\partial}{\partial x_{i+1}}-x_{i+1}\frac{\partial}{\partial x_i}\right)^2
+  E(x_{i},x_{i+1})
\left(x_i\frac{\partial}{\partial x_{i+1}}-x_{i+1}\frac{\partial}{\partial x_i}\right)
\ee
Possible choices for the drift function $E(x,y)$ are:
\ben
\item $E(x,y)=E$ is constant. In that case, the product of Gaussian measures
\eqref{gausseq}
is still invariant if we add
thermalising noise at both ends of the chain at the same temperature $T$. Moreover,
in the case of different temperatures, the equations for the
stationary correlation functions of energies at different sites
are still closed, i.e., the equations for the expectations of polynomials of degree $K$
only involve polynomials of degree $K$. However, duality is lost just as in the 
case of the asymmetric
exclusion process.
\item $E(x,y)=E(x^2+y^2)$. This corresponds to the most general situation
where the product of Gaussian measures is invariant (under the circumstances
of the previous item). In the case of unequal temperatures, equations
for stationary correlations are no longer closed.
\item $E(x,y)=E xy$. This form of the drift has the advantage of
being a product of two $K$-operators in \eqref{Koper}, which is also the case if one
goes from the symmetric to the asymmetric exclusion process (in the quantum spin chain language).
\een

\section*{Acknowledgements}

We thank B. Derrida and E. Presutti for helpful discussions. C.Giardin\`a
acknowledge the kind hospitality of ESPCI and the support of Dutch NWO and 
the program RDSES of European Science Foundation.


\begin{thebibliography}{99}

\bibitem{Mathur1}
L. Balents, M. P. A. Fisher, and M. R. Zirnbauer,
Chiral metal as a ferromagnetic super spin chain,
{\em Nucl. Phys. B} {\bf 483}, 601 (1996).

\bibitem{BF1}
M. Beccaria, C. Presilla, G.F. De Angelis, G. Jona-Lasinio,
An exact represenation of the fermion dynamics in terms of Poisson processes and its connection
with Monte Carlo algorithms, {\em Europhys. Lett.} {\bf 48}  243 (1999).

\bibitem{olla}  C. Bernardin, S. Olla,
Fourier's law for a microscopic model of heat conduction,
{\em J. Stat. Phys.} {\bf 121} 271 (2005).

\bibitem{BGL}
L. Bertini, D. Gabrielli, J.L. Lebowitz,
Large deviations for a stochastic model of heat flow,
{\em J. Stat. Phys.} {\bf 121} 843-885 (2005).

\bibitem{pres}
A. De Masi and E. Presutti, Probability estimates for symmetric simple exclusion random walks.
{\em Ann. Inst. H. Poincar\'{e}} {\bf 19} 71--85 (1983).

\bibitem{DLS} B. Derrida, J.L. Lebowitz, E.R. Speer,
Large deviations of the density profile in the stady state of the
open symmetric simple exclusion process,
{\em J. Stat. Phys.}{\bf 107}  599- (2002).

\bibitem{BF} S.E. Esipov, T,J, Newman,
New formulation of restricted growth process,
{\em J. Stat. Phys.} {\bf 70}  691- (1993).

\bibitem{olla2} J. Fritz, K. Nagy, S. Olla,
Equilibrium Fluctuations for a System of Harmonic Oscillators with Conservative Noise,
{\em J. Stat. Phys.} {\bf 122} 399- (2006).

\bibitem{GK} C. Giardin\`a, J. Kurchan,
The Fourier law in a momentum-conserving chain,
{\em J. Stat. Mech.} P05009 (2005).

\bibitem{landim}
C. Kipnis and C. Landim,
{\em Scaling limits of interacting particle systems},
Springer, Berlin, (1999).

\bibitem{KMP} C. Kipnis, C. Marchioro, E. Presutti,
Heat flow in an exactly solvable model,
{\em J. Stat. Phys.} {bf 27}  65-74 (1982).

\bibitem{coherent1}
J. R. Klauder and B. Skagerstam,
{\em Coherent States Applications in Physics and Mathematical Physics}
World Scientific, Singapore, (1985).

\bibitem{coherent2}
J. Kurchan, P. Leboeuf, and M. Saraceno,
Semiclassical approximations in the coherent-state representation,
{\em Phys. Rev. A} {\bf 40} 6800-6813 (1989).

\bibitem{lao}
C. Landim, A. Milan\'es, S. Olla,
Stationary and Nonequilibrium Fluctuations in Boundary Driven Exclusion Processes,
math.PR/0608165 (2006).

\bibitem{lig} T.M. Liggett, {\em Interacting Particle Systems},
Springer, Berlin, (1985).

\bibitem{coherent}
A. Perelomov, {\em Generalized Coherent States and their
Applications}, Springer, New York, (1986).

\bibitem{schutz-review}G. M. Schutz,
Exactly solvable models for many-body systems far from equilibrium,
in {\em Phase Transitions and Critical Phenomena.} Vol. {\bf 19},
Eds. C. Domb and J. Lebowitz, Academic Press, London, (2000).

\bibitem{SS} G. Schutz and S. Sandow,
Non-Abelian symmetries of stochastic processes: Derivation of correlation functions
for random-vertex models and disordered-interacting-particle systems,
Phys. Rev. {\bf E 49} 2726- (1994).

\bibitem{S} H. Spohn,
Long range correlations for stochastic lattice gases in a non-equilibrium steady state,
{\em J. Phys. A} {\bf 16} 4275-4291 (1983).

\bibitem{stinch} R. Stinchcombe, Stochastic non-equilibrium systems,
{\em Advances in Physics}, {\bf 50} 431-496 (2001).

\bibitem{Mathur} Yi-Kuo Yu, H. Mathur,
Are Directed Waves Multifractal?,
{\em Phys. Rev. Lett.} {\bf 81} 3924- (1998).























\end{thebibliography}
\end{document}